\documentclass[runningheads]{llncs}
\usepackage{graphicx}

\usepackage{tikz}
\usepackage{comment}
\usepackage{amsmath,amssymb} 
\usepackage{color}

\usepackage{times}
\usepackage{epsfig}
\usepackage{graphicx}
\usepackage{amsmath}
\usepackage{amssymb}
\usepackage{bm}
\usepackage{float}
\usepackage{caption}
\usepackage{subcaption}
\usepackage{booktabs}
\usepackage[para,online,flushleft]{threeparttable}
\usepackage[breaklinks=true,bookmarks=false]{hyperref}
\usepackage{xcolor}
\usepackage{mfirstuc}
\usepackage{cleveref}

\begin{document}

\title{Spatiotemporal Feature Learning Based on Two-Step LSTM and Transformer for CT Scans} 

\titlerunning{SFL based on two-step models}
%
\author{Chih-Chung Hsu \and
Chi-Han Tsai \and
Guan-Lin Chen \and
Sin-Di Ma \and
Shen-Chieh Tai}
\authorrunning{C.C. Hsu et al.}
%
\institute{Institute of Data Science, National Cheng Kung University, \\
No.1, University Rd. Tainan City, Taiwan \\
\email{cchsu@gs.ncku.edu.tw}\\
}
\maketitle

\begin{abstract}
Computed tomography (CT) imaging could be very practical for diagnosing various diseases. However, the nature of the CT images is even more diverse since the resolution and number of the slices of a CT scan are determined by the machine and its settings. Conventional deep learning models are hard to tickle such diverse data since the essential requirement of the deep neural network is the consistent shape of the input data. In this paper, we propose a novel, effective, two-step-wise approach to tickle this issue for COVID-19 symptom classification thoroughly. First, the semantic feature embedding of each slice for a CT scan is extracted by conventional backbone networks. Then, we proposed a long short-term memory (LSTM) and Transformer-based sub-network to deal with temporal feature learning, leading to spatiotemporal feature representation learning. In this fashion, the proposed two-step LSTM model could prevent overfitting, as well as increase performance. Comprehensive experiments reveal that the proposed two-step method not only shows excellent performance but also could be compensated for each other. More specifically, the two-step LSTM model has a lower false-negative rate, while the two-step Swin model has a lower false-positive rate. In summary, it is suggested that the model ensemble could be adopted to more stable and promising performance in real-world applications.
\end{abstract}

\section{Introduction}
SARS-CoV-2 (COVID-19), a contagious disease that has caused an extremely high infection rate in the world, can cause many infections in a short time. Doctors need to use the image to confirm the severity of the infection in the lung medical imaging. However, human judgment resources are limited, and a large amount of data is impossibly labeled in a timely manner. Therefore, many studies were proposed to solve the problem of the insufficient workforce via deep learning methods and provide auxiliary systems for preliminary marking or judgment. The lack of a workforce leads to several problems. First, the difference in image resources caused by the types of image settings is mostly obtained by X-ray or computed tomography (CT) imaging schemes, in which the settings of the equipment and resolution are usually diverse, causing poor generalizability of the model. Second, unlike X-ray, CT scan produces more detailed and complete imaging information, so it is possible to see the inside of solid organs. However, rare studies focus on the CT scan due to the high diversity of the CT slices, compared to X-ray-based approaches \cite{hussain2021corodet} \cite{abbas2021classification}\cite{chen2021design}\cite{ismael2021deep}\cite{swin2021Xray}.

There are many real-world challenges in computer vision, aiming to solve the problems of the COVID-19 dataset described above and classify slices. For example, Fang et al.\cite{Conv_Deconv} proposed enhanced local features based on convolution and deconvolution in CT images to improve the classification accuracy. Singh et al.\cite{CNN_MODE} use multi-objective differential evolution–based convolutional neural networks to resolve the difference between infected patients. Pathak et al.\cite{Cla_trans_learning} proposed a transfer learning model to classify COVID-19 patients via the semantic features extracted by ResNet\cite{ResNet}, in which the weights in the ResNet were pretrained by the ImageNet dataset.

In addition, Transformer has significantly contributed to current research in image classification or segmentation. For example, Jiang et al. \cite{swin2021Xray} proposed a method that combined Swin Transformer and Transformer in Transformer module to classify chest X-ray images into three classes: COVID-19, Pneumonia, and Normal (healthy). As reported by \cite{swin2021Xray}, the accuracy could achieve 0.9475. Tuan Le Dinh et al.\cite{GAN} also have done some model experiments with chest X-ray images. They found that the transformer-based models performed better than convolution-based models on all three metrics: precision, recall, and F1-score. However, the datasets were from many open-source datasets.
To some extent, this might affect the model's accuracy. Furthermore, X-ray images obtained from different machines have various image qualities, image color channels as well as resolutions. These factors have a significant impact on the model training pipeline. Although this uses the data proposed by the official competition, which contains a large number of thoracic CT images, the slice positions in the data are different, and the arrows manually marked by the hospital are included, which leads to the high complexity of the data and the difficulty of classification. Third, CT scan results in medical image analysis are usually regarded as three-dimensional data, but In most traditional CNN models, only two-dimensional data operations can be performed, and due to the data set provided by the competition, each case's total number of slices varies.

COV19-CT-DB dataset proposed by Kollias et al.\cite{kollias2022ai}\cite{kollias2021mia}\cite{kollias2020deep}\cite{kollias2020transparent}\cite{kollias2018deep} has large annotated Covid-19 and non-Covid-19 CT scans, which can improve the learning capability for deep learning methods. However, the resolution and length of CT scans in COV19-CT-DB were inconsistent between each CT scan, as described in \cref{fig:dis}; thus, deep neural networks trained by this dataset were difficult to achieve promising results. To deal with this issue, Chen et al.\cite{chen2021adaptive} proposed the 2D and 3D methods to explore the importance of slices of a CT scan. The 2D method, adaptive distribution learning with statistical testing (ADLeaST), combined statistical analysis with deep learning to determine whether the input CT scan is Covid or non-Covid by hypothesis testing. The Covid and non-Covid slices in a CT scan were mapped to follow specific distributions via a deep neural network that the slices without lung tissue can be excluded, thus increasing the stability and explainability of deep learning methods. However, the 2D method could be influenced by implicit slices (i.e., positive slices without obvious symptoms), which were randomly sampled from the Covid CT for training. In the 3D method, a 3-D volume-based CNN (CCAT) was proposed to automatically capture the relationship of slices via the self-attention mechanism in the Within-Slice-Transformer (WST) and Between-Slice-Transformer (BST) modules. However, the 3D methods suffered from a fatal overfitting issue due to the insufficient training samples and the complex architecture of Vision Transformer. 

\begin{figure}[!ht]
	\centering
	\includegraphics[width=0.6\textwidth]{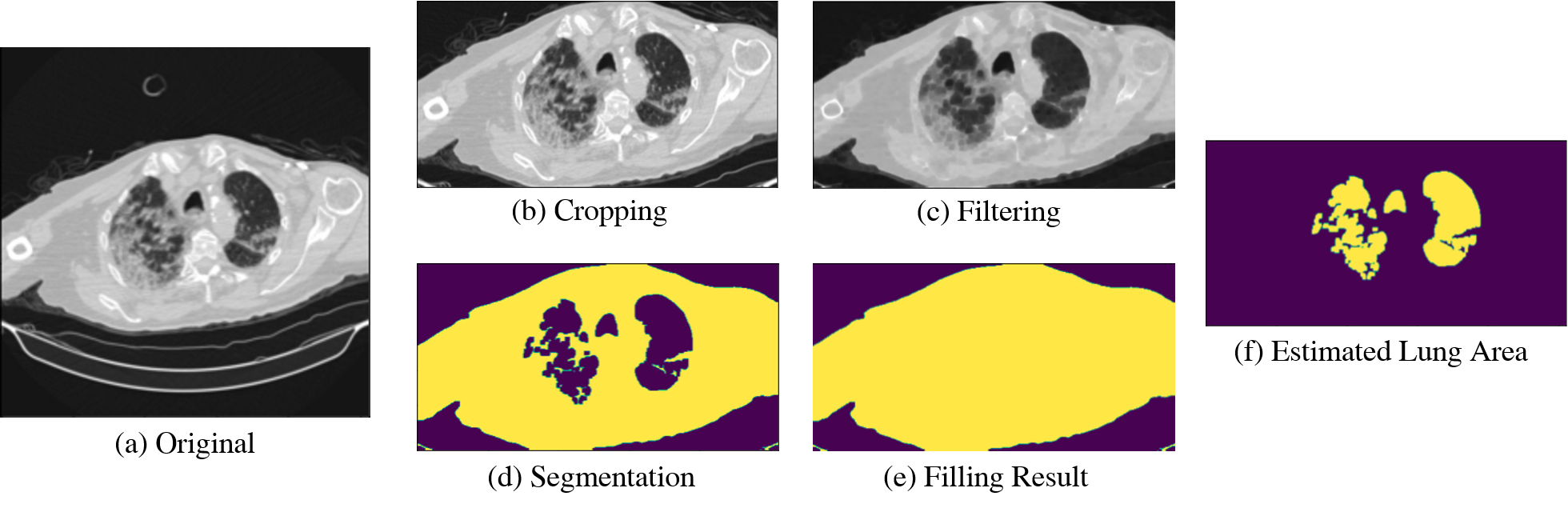}
	\caption{Preprocessing slice for lung area estimating. From (a) to (f) are original slice, cropped slice, filtered slice, segmentation map, filling result obtained by binary dilation algorithm, lung area estimated by the difference between (d) and (e), respectively.}
	\label{fig:lung}	
\end{figure}

Although Chen et al. \cite{chen2021adaptive} made an improvement on the performance by center slice sampling on a CT scan with fixed percentages, we argue that the important slices could also be excluded since the sampling range was imprecise. This research proposed a novel two-step method for the challenge dataset, COV19-CT-DB \cite{kollias2022ai}\cite{kollias2021mia}\cite{kollias2020deep}\cite{kollias2020transparent}\cite{kollias2018deep}, to break the limitation. Firstly, a new strategy based on lung area estimating for slice selection was proposed. In a CT scan, the slice with a larger area of lung tissue is more likely to contain symptoms of Covid, since lungs are the organs most affected by Covid‐19. The 2D model will be trained by the slices consisted with the most lung tissue, thus making the embedding features more representative. Secondly, the embedding features from the 2D model pretrained on the selected slice are used to explore the relationship between slices in a CT scan via the two-step models. The two-step models considered the temporal variation of a CT scan for classifying without limited range of slices.

\section{Methodology}

In this paper, we mainly focus on improving the 2D CNN model by the enhanced strategy for selecting a range of slices . The selected slices will be different for each CT scan, thus improving the training quality to get a better result. In addition, the embedding features from a well trained 2D model was utilized to automatically explore the relevance of each slice. The slice selection algorithm and two-step models are introduced in \Cref{sec:lung} and \Cref{sec:swinv2+embedding}, respectively.

\subsection{Lung Area Estimating for Slice Selection}
\label{sec:lung}
Given a CT scan $C=\{\mathbf{X}_i\}^{n}_{i=1}$ with label $y$ and $n$ slices, we aim to discover the range of slices $C^*=\{\mathbf{X}_i\}^{e}_{i=s}$ which contains the largest area of lung tissue with start index $s$ and end index $e$. Firstly, a cropping algorithm was employed to exclude the background of a CT scan slice. The cropped region $C_{crop}=\{(\mathbf{X}_{crop}^{(i)})\}^{e}_{i=s}$ can reduce the computational complexity and increase the estimating precision for all slices in a CT scan, as shown in \cref{fig:lung}(b). Then a filtering operator is adopted to eliminate the noise caused by the CT scanner, which can be defined as:
\begin{equation}
    \label{eq:filter}
    \mathbf{X}_{filter} = Filter(\mathbf{X}_{crop}).
\end{equation}
Afterwards, the segmentation map $\mathbf{M}$ of the filtering slice can be determined by a specific threshold $t$, as shown in \cref{fig:lung}(d). The process can be written as:
\begin{equation}
    \label{eq:seg}
    \mathbf{M}[i,j] = 
    \begin{cases}
    0,\,\text{if}\,\mathbf{X}_{filter}[i,j] < t\\
    1,\,\text{if}\,\mathbf{X}_{filter}[i,j] >= t
    \end{cases}
\end{equation}
where $i$ is the width index and $j$ is the height index of a slice. In order to discover the lung area, the binary dilation algorithm \cite{enwiki:1082436538} was adopted to calculate the filling result $\mathbf{M}_{filling}$. The lung region can be obtained by the difference of the segmentation map and its filling map, as illustrated in \cref{fig:lung}(d)(e)(f). Finally, the lung area of a slice can be defined as:
\begin{equation}
    \label{eq:area}
    Area(\mathbf{X}) = \sum_i\sum_j\mathbf{M}_{filling}(i,j) - \mathbf{M}(i,j)
\end{equation}
The start index $s$ and end index $e$ can be determined by the accumulating area, which can be formulated as:
\begin{equation}
    \label{eq:area}
    \begin{split}
    \mathop{\arg\max}_{s,\,e}{\sum^e_{i=s}Area(\mathbf{X}_i)}, \\
    \text{subject to } e-s \leq n_c,
    \end{split}
\end{equation}
where $n_c$ is the number of sampling slices.

After calculating the slice range $C^*$, we use the sub-dataset as the training data to train the 2D model by the binary cross-entropy loss, which can be defined as:
\begin{equation}
    \label{eq:bce}
    L_{bce} = -\frac{1}{N}\sum^N_{i=1}y_i\cdot\log(p_i)+(1-y_i)\log(1-p_i),
\end{equation}
where $p_i=F_{2d}(\mathbf{X}_i;\theta_{2d})$, $\theta_{2d}$ is the trainable parameters of the 2D model, and $N$ is the total number of selected slices.

\subsection{Two-Step models}
\label{sec:swinv2+embedding}
For the Covid-19 images classification, in order not to be limited to the extracted slices for prediction, we use the temporal model to automatically learn the relevance of the embedding features from the 2D model. Then we will describe the model selected in this paper as lstm and swin, and describe the reasons and its advantages and disadvantages.

\subsubsection{LSTM}
Long Short Term Memory(LSTM) Network is an advanced RNN, a sequential network, that allows information to persist. It is capable of handling the vanishing gradient problem faced by RNN. A recurrent neural network is also known as RNN is used for persistent memory. Similarly RNNs work, they remember the previous information and use it for processing the current input. The shortcoming of RNN is, they can not remember Long term dependencies due to vanishing gradient. LSTM are explicitly designed to avoid long-term dependency problems. On the other hand, LSTM are prone to overfitting and it is difficult to apply the dropout algorithm to curb this issue. We have tried to use the LSTM model to learn the difference between Covid-19 and non-Covid-19 CT scans with the embedding features, which were randomly selected and sorted. In other words, we have trained the LSTM model by dropping out the embedding of slices to solve the overfitting problem and keep consistent information of the slices simultaneously..

\subsubsection{Swin Transformer}
Since Swin Transformer was proposed by Liu et al.\cite{swin}, this model architecture had served as a general-purpose backbone for computer vision. They had solved the challenge, adapting Transformer from language to vision arise from differences between the two domains, such as large variations in the scale of visual entities and the high resolution of pixels in images compared to words in text.
Recently, Swin Transformer V2 was updated by Liu et al.\cite{swin_v2}, setting new performance records on 4 representative vision tasks, including ImageNet-V2 image classification, COCO object detection, ADE20K semantic segmentation, and Kinetics-400 video action classification.

\section{Experiments}
\subsection{Experimental Settings}

\subsubsection{Implementation details.}
The challenge dataset, COV19-CT-DB \cite{kollias2022ai}\cite{kollias2021mia}\cite{kollias2020deep}\cite{kollias2020transparent}\cite{kollias2018deep} comprises 1993, 484, 5281 CT scans for training, validation, and testing, respectively. We found that the size of slices is inconsistent in some CT scans that only the last slice is scanned from the common axis (i.e., horizontal plane). To deal with this issue, we manually removed the CT scans with an uncommon scanned axis for the training dataset (less than 1\%). In addition, the last slice in these cases was selected for both training and testing. The rest of the CT scans were selected by the proposed method as described in \Cref{sec:lung}. The sampling size of slices was set to half of the total sizes for each CT scan.

Sensitivity and Specificity play vital roles in evaluating the accuracy and reliability of a diagnostic test both prior to implementation and as part of ongoing quality assurance. \textbf{Sensitivity} relates to how well a test can detect the presence of COVID-19 disease (the percentage of true positive results in patients who have the disease). Higher test sensitivity equates to positive infection and means there is a lower rate of false-negative results.\textbf{Specificity} relates to how well a test can confirm the absence of COVID-19 infection. It indicates the percentage of true negative results in patients who don't have the disease. Higher Specificity will mean a lower rate of false-positive results for the test.

\subsubsection{2D CNN model.}

The EfficientNet-b3a \cite{rw2019timm} was adopted as the backbone for our 2D CNN model, and we add an fully connected layer with embedding size 224 to reduce the computational complexity for the proposed two-step model, which explored the embedding features. For training, sixteen slices were randomly sampled from the selected slices as a batch. Adam \cite{Adam} was used as the optimizer with learning rate 0.0001, $\beta_1=0.9$, $\beta_2=0.999$ and the weight decay is 0.0005. For testing, the average of the predicted probabilities of sixteen slices sampled from a CT scan was used to judge the result, where the threshold was set to 0.5.

\subsubsection{LSTM model.}
\label{sec:LSTM model}
We use the weights of the 2d cnn to get the 224 size embedding of each slice, and randomly selected 120 slices embedding of a CT scan as an lstm model input. We used a lstm model have 4 bi-lstm layers with 128 hidden size, and have a linear layer with 64 size before output.

\subsubsection{Swin model.}
Same as LSTM model, we got the 224 size embedding of each slice. Due to the number of slices of CT scans is not equal, we used a random sampling strategy to select 224 slices, for representing one single CT scan. If the number of slices of one single CT scan is less than 224, we select all slices, and the remaining slices are randomly selected from all slices; otherwise, we random sample 224 slices without replacement. 

\begin{table}[!ht]
    \centering
    \caption{Performance evaluation of validation set of the proposed methods and other peer methods in terms sensitivity (SE), specificity (SP), and macro-F1-score. (Mean 10) denotes that the result is obtained by the the average of predictions which came from ten different sampled set of slices.}
    \begin{tabular}{l c c c c} 
      \toprule 
       & SE & SP & Macro-F1 \\
      \midrule
      ADLeasT \cite{chen2021adaptive}     & 0.833 & 0.978 & 0.911 \\
      \midrule
      2D CNN     & 0.897 & 0.959 & 0.930   \\
      2D CNN (Mean 10)              & 0.893 & 0.973 & 0.936   \\
      2D CNN (Mean 50)              & 0.893 & 0.974 & \textbf{0.937}   \\
      \midrule
      2-step LSTM    & \textbf{0.900} & 0.959 & 0.932   \\
      2-step Swin    & 0.879 & \textbf{0.981} & 0.934 \\ 
      \bottomrule
    \end{tabular}
    \label{tab:result1}
\end{table}

\subsection{Performance evaluation}

The experiment result is shown in \cref{tab:result1}. On the 2D CNN model, our f1 score is 0.930 that higher than ADLeasT 0.911 f1 score. However, the f1 score of the 2D CNN model will be slightly different(between 0.920 and 0.938), which is caused by randomly selecting 16 slices from each CT scan. So, we use an in-model ensemble, averaging our predictions for each CT scan to get more robust predictions. After 50 times averaging, the f1 score of the 2D CNN model becomes 0.937. In fact, it is impractical to use this method because it will take a lot of time. 
%

\section{Conclusions}

In this paper, we proposed a method to effectively and efficiently resolve the problems in COV19-CT-DB \cite{kollias2022ai}\cite{kollias2021mia}\cite{kollias2020deep}\cite{kollias2020transparent}\cite{kollias2018deep} raised by the inconsistent resolution and length of CT scans between different CT scans. By randomly selecting slices from each CT scan, avoid not only 2-step LSTM model overfitting but also increase the performance of the 2D CNN model. Furthermore, we found a 2-step method based on the 2D CNN model, having two higher and different performances on the validation sets. The 2-step LSTM model has a lower false-negative rate, and the 2-step Swin model has a lower false-positive rate.

%
%
\bibliographystyle{splncs04}
\bibliography{egbib}
\end{document}